\documentclass[sigconf]{acmart}
\usepackage{tabularx}
\usepackage{adjustbox}
\usepackage{multicol}
\usepackage{rotating}
\usepackage{hyperref}
\usepackage{multirow}
\usepackage{pifont}
\usepackage{graphicx}
\usepackage{booktabs}
\usepackage{graphics}
\usepackage{xspace} 
\usepackage{color,soul}
\usepackage{booktabs}
\usepackage{mdwlist}
\usepackage{multirow}
\usepackage{amsmath}
\usepackage{enumitem} 
\usepackage{calc}
\usepackage{colortbl}
\usepackage[ruled,linesnumbered]{algorithm2e}
\usepackage{booktabs}
\usepackage{mathtools}
\usepackage{balance}
\usepackage{comment}
\usepackage{textcomp}
\usepackage{subcaption}
\usepackage{tcolorbox}
\usepackage{listings}
\usepackage{xcolor}
\usepackage{colortbl}

\definecolor{magicmint}{rgb}{0.67, 0.94, 0.82}
\definecolor{melon}{rgb}{0.99, 0.74, 0.71}

\newcommand{\nores}{\cellcolor{lightgray}}

\AtBeginDocument{%
  }

\copyrightyear{2024}
\acmYear{2024}
\setcopyright{rightsretained}
\acmConference[JCDL '24]{The 2024 ACM/IEEE Joint Conference on Digital Libraries}{December 16--20, 2024}{Hong Kong, China}
\acmBooktitle{The 2024 ACM/IEEE Joint Conference on Digital Libraries (JCDL '24), December 16--20, 2024, Hong Kong, China}\acmDOI{10.1145/3677389.3702593}
\acmISBN{979-8-4007-1093-3/24/12}

\begin{document}

%
\title[Unveiling Temporal Trends in 19th Century Literature]{Unveiling Temporal Trends in 19th Century Literature: An Information Retrieval Approach}

\author{Suchana Datta}
\affiliation{%
  \institution{University College Dublin, Ireland}
  \country{}
}
\email{suchana.datta@ucd.ie}
\author{Dwaipayan Roy}
\affiliation{%
  \institution{IISER Kolkata, India}
  \country{}
}
\email{dwaipayan.roy@iiserkol.ac.in}
\author{Derek Greene}
\affiliation{%
  \institution{University College Dublin, Ireland}
  \country{}
}
\email{derek.greene@ucd.ie}
\author{Gerardine Meaney}
\affiliation{%
  \institution{University College Dublin, Ireland}
  \country{}
}
\email{gerardine.meaney@ucd.ie}

\renewcommand{\shortauthors}{Datta et al.}

\begin{abstract}

In English literature, the 19th century witnessed a significant transition in styles, themes, and genres. Consequently, the novels from this period display remarkable diversity. This paper explores these variations by examining the evolution of term usage in 19th century English novels through the lens of information retrieval.
By applying a query expansion-based approach to a decade-segmented collection of fiction from the British Library, we examine how related terms vary over time.
Our analysis employs multiple standard metrics including Kendall's tau, Jaccard similarity, and Jensen-Shannon divergence to assess overlaps and shifts in expanded query term sets.
Our results indicate a significant degree of divergence in the related terms across decades as selected by the query expansion technique, suggesting substantial linguistic and conceptual changes throughout the 19th century novels.

\end{abstract}




\maketitle

\section{Introduction and background} \label{sec:intro}


The 19th century was a period covering a profound transition in literature markets with the emergence of new literary forms, styles and thematic varieties.
From the \emph{Romanticism} and \emph{Fictions} of the early 1800s to the \emph{Realism} and \emph{Naturalism} of the later half of the century, the novels written in this era mirror the complex and fascinating socio-political and cultural changes that took place. 
Influenced by several historical contexts, literature movements and the changing tastes of readers, the language and vocabulary used in the literature also evolved over this period.


In the field of information retrieval (IR), understanding how terms evolved over time is critical when developing search systems that can effectively retrieve relevant and useful information from historical collections.
For traditional search systems, based on term overlap, one of the most challenging issues faced by researchers is the \emph{vocabulary mismatch}.
This arises when the terms used in a search session differs from those terms present in the relevant documents, which leads to poor retrieval performance.
This issue can occur because different individuals may use various words to convey the same or similar concept, leading to low term overlap during retrieval.
Query expansion, a widely-used technique in IR, aims to mitigate this problem by adding semantically similar terms to the search query~\cite{AZAD20191698}.
This is an active research area in IR, where various models have been proposed to address the vocabulary mismatch problem~\cite{10.1145/2348283.2348354,10.1145/3624988,10.1145/3626772.3657979,10.1145/3626772.3661367}.
A substantial body of research has investigated the impact of temporal signals on search \cite{10.1145/1458082.1458320,10.1145/2009916.2009984,10.1145/956863.956951,10.1145/2600428.2609575,10.5555/2964060.2964106}. From these studies, it is well established that the recency of events, in general, improve modelling of the temporal aspects of queries and documents can enhance retrieval effectiveness.
However, the problem remains a concern when applied to text from different time periods.
This is due to the ever-evolving landscape of literature, where the usage of words changes significantly over time. This means that a term relevant in one decade may not carry the same importance in another, and may eventually become extinct.

Previously, a range of approaches have been applied to examine trends in word usage in large text corpora. Simple methods include word frequency analysis~\cite{bybee2006frequency} and examining the appearance of Ngrams over time (e.g. the Google Books Ngram Viewer). However, such tools need to be treated carefully, depending on the composition of the corpus being studied and due to the shifts in word meanings over time \cite{pettit2016historical}. To identify such shifts as a means of investigating cultural phenomena, authors have examined distributional similarity models~\cite{gulordava2011distributional}. Considerable work has involved the generation of diachronic word embeddings~\cite{hamilton2016diachronic,kutuzov2018diachronic}. Given that word embeddings can capture word meaning in a static corpus, training embeddings on different time periods can allow meaning representations at different time periods to be compared, supporting the identification of changing contexts around particular words of interest. However, this requires sufficient data from each time period to construct a robust embedding.

This study aims to explore trends in over $10,000$ 19th century novels 
from
the British Library Digital Collection (BL19)\footnote{See \url{https://labs.biblios.tech/}}. 
By splitting the collection of novels into decades, we investigate how the expansion terms selected through the Relevance-based Language Model (RLM)~\cite{10.1145/383952.383972,DBLP:conf/trec/JaleelACDLLSW04} for the same query differ from one decade to another.
Specifically, we analyse how the selected expansion terms and their associated weights vary when RLM is applied to novels published in different decades, as well as when applied to the entire collection of novels.
Our hypothesis is that the expansion terms with the highest weights will exhibit significant variations across different decades, reflecting the evolving nature of the language.
Through this study, we uncover nuanced differences in the semantic characteristics of the language by comparing the expansion terms generated from each decade with the same produced from the entire collection taken together. 
The primary contributions 
of this work 
are as follows:
\begin{itemize}
[leftmargin=*]
    \item We conduct an in-depth analysis of how specific concepts evolved in the BL19 fiction collection, based on semantically similar terms identified by the pseudo-relevance feedback. Specifically, we measure the correlation between the expansion terms estimated via the relevance feedback model for a given decade in the 19th century and those from the full collection. This allows us to gain an insight into how the terms used in novels evolved over time.
    \item We demonstrate that applying relevance feedback on a given decade from the 19th century exhibits different relevance behaviour in comparison to feedback from the entire time span of the collection.
\end{itemize}




\section{Quantifying the evolution of concepts}\label{sec:method}
In this paper, we aim to capture how the relevance and meaning of key concepts have changed over different decades by analyzing term overlaps, rank correlations, and semantic shifts for understanding the temporal dynamics of language and themes in literary texts.
The comparison is structured around several key analyses that collectively offer insights into understanding how the usage of words has evolved resulting in varying term selection, retrieval outcomes, and semantic meanings across different periods in the collection.
First, the analysis examines the overlap of the most significant expansion terms selected by RLM from different sub-collections. 
We can identify the degree of consistency in term selection across periods by comparing the top terms generated for each decade.
Additionally, the overlap in terms selected across decades is quantified by calculating the Jaccard similarity between feedback queries generated from different sub-collections.
This serves to highlight which terms maintain their relevance over multiple decades and which ones are unique to specific periods.
Another approach is to measure the similarity between ranked lists produced by queries expanded from the entire collection and those expanded from individual sub-collections (decades).
Using Kendall's $\tau$ to quantify rank correlation, this comparison provides insights into how closely the retrieval outcomes from sub-collections align with those from the full collection.
The differences in term weight distributions across decades are further explored using Jensen-Shannon (JS) divergence. 
%
Through these analyses, we aim to provide a comprehensive view of the temporal evolution of language and themes in literary texts, highlighting some valuable insights into how the concepts have evolved, across different historical decades in the collection.

\begin{table}[!t]
\centering
\small
\caption{
\small
Number of novels in each decade in BL19 collection.}
\begin{adjustbox}{width=0.85\columnwidth}
\begin{tabularx}{\columnwidth}{lccccccc}
\toprule
Decades & 1830s & 1840s & 1850s & 1860s &1870s & 1880s & 1890s\\
\cmidrule{2-8}
\#Novels & 20 & 144 & 746 & 1139 & 1750 & 2034 & 4377 \\
\bottomrule

\end{tabularx}
\label{tab:coll}
\end{adjustbox}
\end{table}
\begin{table}[!t]
\centering
\small
\caption{
\small
Query terms of three different categories that are used to study the temporal changes in the BL19 collection.}
\begin{adjustbox}{width=0.9\columnwidth}
\begin{tabularx}{\columnwidth}{lX}

\toprule
Thematic & immigrant, emigrant, foreign, newcomer, alien, enslaved, colony, vampire \\
\cmidrule{2-2}
Plot & engagement, proposal, wedding, suitor, lover, betrothal, eligible, consent, love, mesalliance, heiress, eviction \\
\cmidrule{2-2}
Genre & crime, murder, mystery, villain, adventure \\
\bottomrule

\end{tabularx}
\label{tab:qterms}
\end{adjustbox}
\end{table}

\section{Experimental setup}

\paragraph{\textbf{Collection}} In our experiments, we focus on the set of $10,210$ English-language fiction texts from the BL19 collection.
Their publication dates span from the 1830s to the end of the 19th century, representing a key period both historically and in terms of the development of literary fiction.
These texts include well-known novels by prominent authors such as Charles Dickens and Jane Austen and a significant number of little-known and infrequently studied works.
As such, this potentially provides a mirror on attitudes and views towards important societal issues during this time period, such as gender, migration, and public health. Furthermore, it presents an opportunity to understand how word usage and meanings have changed around these topics during the course of the 19th century. Studying semantic change and word usage in this way allows humanities researchers to understand the evolution of language and how meanings of words have shifted over time, reflecting cultural and societal changes.
These texts were grouped into seven decades, spanning from the 1830s to the 1890s. The distribution of texts across these decades is uneven, with a notable increase in the late 19th century (see Table \ref{tab:coll}). This pattern mirrors both the composition of the BL19 collection and the broader rise in British fiction publication during that period.
Henceforth, we will refer to the entire collection by $\mathcal{C}$.
Further, the sub-collections consisting of novels published in the 1830s, 1840s, and subsequent decades will be denoted by $\mathcal{C}^{30}$, $\mathcal{C}^{40}$, and so on.
More generally, let $\mathcal{C}^i$ denote any sub-collection corresponding to a specific decade $i$.
For our experiments, we split each sub-collection $\mathcal{C}^i$ in the unit of paragraphs prior to retrieval. 

\begin{table*}[!t]

\centering

\caption{
\small
Top 15 feedback terms chosen by RLM for two sample queries (`immigrant’ and `murder’) from Table \ref{tab:qterms}. Feedback terms are obtained from different decades and also from the entire 19th century fictions (see the last row of each group). Term overlaps between two decades are in bold and the decade in which the corresponding query term does not occur is grayed out.
}
\label{tab:sample}

\begin{adjustbox}{width=0.82\textwidth}
\begin{tabularx}{\textwidth}{llX}
\toprule

Query & Time Period & Feedback terms\\

\midrule 

& 1830-40
& \nores \\

\cmidrule{3-3}

\multirow{8}{*}{\begin{turn}{90}Immigrant\end{turn}}
& 1841-50
& pudent, avez, censu, raison, soulless, asop, axl, aptli, diadem, altitud, radic, fry, shred, wight, cape\\

\cmidrule{3-3}

& 1851-60
& trade, \textbf{countri}, emigr, soil, \textbf{saxon}, \textbf{great}, flourish, melbourn, cite, irish, gener, furrugn, \textbf{nativ}, anccstii, intrus\\

\cmidrule{3-3}

& 1861-70
& britain, \textbf{land}, foreign, \textbf{govern}, race, distdleri, \textbf{saxon}, \textbf{countri}, law, influx, \textbf{great}, thed, tax, compani, natur\\

\cmidrule{3-3}

& 1871-80
& \textbf{coloni}, \textbf{govern}, \textbf{land}, \textbf{countri}, peopl, \textbf{home}, durban, \textbf{australia}, popul, state, settler, dimsdal, \textbf{good}, number, \textbf{nativ}\\

\cmidrule{3-3}

& 1881-90
& \textbf{land}, \textbf{countri}, queensland, room, \textbf{nativ}, work, men, \textbf{australia}, \textbf{home}, \textbf{great}, \textbf{year}, place, chines, trial, \textbf{time}\\

\cmidrule{3-3}

& 1891-99
& \textbf{alien}, \textbf{favish}, \textbf{land}, \textbf{time}, young, turn, yenta, \textbf{chananya}, \textbf{year}, man, mendel, dai, america, \textbf{good}, found\\

\cmidrule{2-3}

& \textbf{1831-99}
& \textbf{alien}, \textbf{coloni}, \textbf{favish}, \textbf{countri}, \textbf{land}, \textbf{turn}, emigr, \textbf{chananya}, \textbf{year}, \textbf{found}, \textbf{time}, \textbf{good}, make, \textbf{australia}, thing\\

\midrule
\midrule

& 1831-40
& gipsei, girl, walsingham, oliv, \textbf{blood}, clifford, \textbf{man}, \textbf{deed}, \textbf{exclaim}, twist, \textbf{time}, fear, \textbf{heard}, moment, repli\\

\cmidrule{3-3}

\multirow{8}{*}{\begin{turn}{90}Murder\end{turn}}
& 1841-50
& \textbf{man}, \textbf{commit}, \textbf{bodi}, \textbf{cri}, \textbf{crime}, \textbf{guilti}, \textbf{found}, \textbf{kill}, \textbf{evid}, \textbf{case}, \textbf{accus}, person, \textbf{heard}, \textbf{head}, \textbf{deed} \\

\cmidrule{3-3}

& 1851-60
& \textbf{commit}, \textbf{man}, \textbf{deed}, \textbf{crime}, \textbf{cri}, \textbf{accus}, \textbf{death}, \textbf{word}, \textbf{blood}, answer, \textbf{exclaim}, \textbf{case}, \textbf{bodi}, \textbf{head}, \textbf{evid}\\

\cmidrule{3-3}

& 1861-70
& \textbf{man}, \textbf{commit}, \textbf{crime}, \textbf{bodi}, \textbf{dead}, \textbf{heard}, \textbf{cri}, \textbf{word}, \textbf{mr}, \textbf{blood}, men, \textbf{horror}, prove, \textbf{found}, \textbf{hear}\\

\cmidrule{3-3}

& 1871-80
& \textbf{commit}, \textbf{man}, \textbf{blood}, \textbf{death}, \textbf{bodi}, \textbf{heard}, finn, \textbf{crime}, \textbf{found}, night, \textbf{mr}, bonteen, \textbf{guilti}, \textbf{evid}, phinea\\

\cmidrule{3-3}

& 1881-90
& \textbf{commit}, \textbf{man}, \textbf{crime}, victim, \textbf{polic}, \textbf{kill}, \textbf{found}, \textbf{case}, mysteri, arrest, \textbf{hand}, \textbf{mr}, dai, \textbf{blood}, \textbf{time}\\

\cmidrule{3-3}

& 1891-99
& \textbf{man}, \textbf{commit}, \textbf{crime}, \textbf{dead}, \textbf{polic}, \textbf{kill}, \textbf{horror}, face, \textbf{bodi}, peopl, \textbf{word}, \textbf{blood}, \textbf{found}, \textbf{hand}, \textbf{cri}\\

\cmidrule{2-3}

& \textbf{1831-99}
& \textbf{commit}, \textbf{man}, \textbf{horror}, \textbf{kill}, \textbf{dead}, \textbf{crime}, \textbf{cri}, \textbf{victim}, \textbf{bodi}, \textbf{blood}, \textbf{hand}, \textbf{death}, call, life, \textbf{hear}\\

\bottomrule

\end{tabularx}
\end{adjustbox}
\end{table*}

\paragraph{\textbf{Topics}} We consider a diverse list of $25$ query keywords from three distinct categories (thematic, plot, and genre), as identified by experts of 19th century British and Irish literature (see Table \ref{tab:qterms}). Many of these keywords broadly reflect key areas of societal change during the 19th century, including gender and migration. 
Thematic terms associated with 19th century migration could be expected to occur more frequently as both inward and outward migration increased exponentially from Victorian Britain\footnote{In British history, the Victorian era is the period between approximately 1820 and 1914, corresponding roughly but not exactly to the period of Queen Victoria’s reign.}. The frequency of terms relating to it offers insight into the relationship between the fiction in BL19 and historical developments. However, the fact that the lexical search indicates a strong association between the terms `immigrant’ and `vampire’ would indicate that this topic is not treated with sober realism in this body of fiction and that fear and suspicion dominate.
`Engagements' and `proposals' are key elements in both the domestic, realist fiction which came to dominate fiction from the early to the mid nineteenth century. The relative frequency of terms associated with this marriage plot as the century progressed offers an insight into the extent to which this form of fiction dominates the corpus. It also indicates the extent to which plots and issues primarily concerned with the life of young women are a central focus of that fiction over the period of analysis.

In contrast, terms associated with `mystery' and `crime' in BL19 indicate the extent to which the relatively new genre of crime fiction is represented. Modern accounts identify the publication of short stories of Edgar Allan Poe in the 1840s and the novels of Arthur Conan Doyle in the 1880s as key points in the development of the genre. The frequency of terms characteristic of this genre offers an opportunity to test these dates against this corpus, though the fact that the lexical search turned up adventure as a closely associated term indicates that genre boundaries are not rigid in
this period.



\begin{table*}[!t]

\centering

\caption{
\small
Comparisons of rank correlation values (measured with Kendall's $\tau$) between two re-ranked lists retrieved by RLM augmented queries, where one query is obtained from the entire century collection and the other is estimated from an individual decade (see the leftmost column). For each query, the correlations are to be compared between decades, i.e. along the columns, and the highest and lowest correlations are respectively bold-faced and underlined. Grey cells indicate the absence of the query term in the corresponding decade's collection.
}
\label{tab:kendall}

\begin{adjustbox}{width=1\textwidth}

\begin{tabular}{@{}lrrrrrrrrrrrrrrrrrrrrrrrrr@{}}

\toprule

& \begin{turn}{60}Immigrant\end{turn}
& \begin{turn}{60}Emigrant\end{turn}
& \begin{turn}{60}Foreign\end{turn}
& \begin{turn}{60}Newcomer\end{turn}
& \begin{turn}{60}Alien\end{turn}
& \begin{turn}{60}Enslaved\end{turn}
& \begin{turn}{60}Colony\end{turn}
& \begin{turn}{60}Vampire\end{turn}
& \begin{turn}{60}Engagement\end{turn}
& \begin{turn}{60}Proposal\end{turn}
& \begin{turn}{60}Wedding\end{turn}
& \begin{turn}{60}Suitor\end{turn}
& \begin{turn}{60}Lover\end{turn}
& \begin{turn}{60}Betrothal\end{turn}
& \begin{turn}{60}Eligible\end{turn}
& \begin{turn}{60}Consent\end{turn}
& \begin{turn}{60}Love\end{turn}
& \begin{turn}{60}Mesalliance\end{turn}
& \begin{turn}{60}Heiress\end{turn}
& \begin{turn}{60}Crime\end{turn}
& \begin{turn}{60}Murder\end{turn}
& \begin{turn}{60}Mystery\end{turn}
& \begin{turn}{60}Villain\end{turn}
& \begin{turn}{60}Adventure\end{turn}
& \begin{turn}{60}Eviction\end{turn} \\

\cmidrule{2-26}

1830s
& \nores & 0.069 & 0.010 & -0.115 & -0.041
& -0.049 & -0.021 & \nores & 0.013 & -0.004
& \underline{-0.031} & 0.011 & 0.007 & -0.038 & -0.004
& 0.042 & 0.019 & \nores & -0.009 & \textbf{0.122}
& \textbf{0.103} & -0.005 & \underline{-0.022} & -0.031 & \nores \\

1840s
& \underline{-0.024} & \underline{-0.059} & \textbf{0.076} & -0.017 & 0.008
& -0.047 & 0.008 & 0.027 & -0.017 & 0.032
& 0.009 & \textbf{0.043} & 0.012 & -0.020 & 0.031
& -0.013 & -0.007 & 0.024 & 0.021 & 0.042
& 0.018 & \underline{-0.022} & 0.019 & \textbf{0.039} & -0.021 \\

1850s
& -0.021 & -0.082 & -0.076 & \underline{-0.033} & -0.139
& 0.031 & 0.007 & 0.015 & -0.029 & 0.021
& 0.019 & -0.009 & \textbf{0.073} & -0.037 & \underline{-0.028}
& -0.011 & 0.028 & \textbf{0.041} & 0.028 & \underline{0.009}
& 0.037 & 0.028 & 0.049 & \underline{-0.052} & 0.019 \\

1860s
& 0.059 & -0.002 & 0.041 & 0.095 & \textbf{0.113}
& \textbf{0.076} & -0.013 & 0.023 & 0.034 & \underline{-0.021}
& 0.045 & 0.038 & -0.005 & 0.014 & 0.056
& 0.011 & \textbf{0.078} & \underline{-0.023} & \underline{-0.075} & 0.089
& 0.023 & \textbf{0.063} & -0.008 & 0.026 & 0.042 \\

1870s
& -0.002 & 0.091 & \underline{-0.078} & -0.003 & \underline{-0.084}
& \underline{-0.061} & \underline{-0.043} & \textbf{0.029} & \textbf{0.052} & \textbf{0.090}
& -0.007 & -0.024 & 0.038 & \textbf{0.051} & 0.032
& \underline{-0.023} & -0.006 & 0.013 & -0.054 & 0.040
& 0.022 & 0.018 & 0.021 & 0.009 & \textbf{0.065} \\

1880s
& -0.017 & \textbf{0.130} & 0.043 & 0.059 & 0.028
& 0.065 & 0.028 & -0.021 & -0.030 & 0.041
& 0.033 & 0.013 & \underline{-0.014} & \underline{-0.047} & 0.029
& \textbf{0.046} & 0.038 & -0.009 & 0.011 & 0.015
& 0.039 & 0.048 & \textbf{0.081} & -0.039 & 0.041 \\

1890s
& \textbf{0.025} & -0.013 & 0.031 & \textbf{0.107} & 0.087
& 0.039 & \textbf{0.081} & \underline{-0.023} & \underline{-0.041} & 0.007
& \textbf{0.056} & \underline{-0.042} & 0.072 & -0.029 & \textbf{0.081}
& -0.021 & \underline{-0.049} & 0.032 & \textbf{0.051} & 0.027
& \underline{0.009} & 0.054 & 0.021 & -0.042 & \underline{-0.037} \\

\bottomrule

\end{tabular}
\end{adjustbox}
\end{table*}

\section{Results and analysis} \label{sec:res}


We explore whether the demarcation of the collection by decade leads to different outcomes by comparing the two expanded queries generated from the individual sub-collections based on these time periods. 
Specifically, we form expanded queries $EQ_C$ and $EQ_C^i$ using RLM respectively from $\mathcal{C}$ and $\mathcal{C}^i$ and perform a retrieval on the entire collection $\mathcal{C}$. 
We apply BM25~\cite{croft_poisson,bm25_beyond} both for the initial and second stage retrieval.
Specifically, we choose top-scored $100$ expansion terms that the RLM estimates from the initial set of $100$ documents retrieved by BM25.  
In Table~\ref{tab:sample}, we present the top 15 terms with the highest weights as determined by RLM for the queries `immigrant’ and `murder’, generated from both the individual sub-collections based on decades and the entire fiction collection. 
The terms that appear consistently across different decades are highlighted in bold.
From Table~\ref{tab:sample}, we observe the terms contextually similar to `immigrant’ exhibit less overlap across decades in comparison to those related to `murder’.
This suggests that the semantic context of the term `immigrant’ may have varied more significantly over time, while the context for `murder’ remained more stable.
One interesting observation is the presence of `Australia' among the top terms. 
Note that, the late 18th and early 19th centuries marked a pivotal turning point in Australian history, as this period witnessed the establishment of British colonial rule in Australia.
The arrival of the First Fleet in 1787-1788 ushered in a new era, transforming the landscape, society, and culture of this continent through the introduction of European systems, values, and languages\footnote{\url{https://en.wikipedia.org/wiki/Immigration_history_of_Australia} and \url{https://www.homeaffairs.gov.au/about-us-subsite/files/immigration-history.pdf}.}.
From the results, we can see that this historical backdrop has also been reflected in British novels of the time.
The themes, vocabulary, and narratives found in these literary works echo the broader changes occurring within the British Empire, including its colonial expansion and the resulting cultural exchanges. 
This suggests a literary consciousness in British novels that parallels the historical events of the period, capturing the impact of colonialism on both the colonized and the colonizers.
Another interesting term is `alien', rooted in the Latin word \emph{alienus} meaning ``belonging to another''.
Although the \emph{Aliens Act} was imposed by the British Parliament in 1793, the use of the term associated with immigrants in novels appeared in the late 1800s.

The impact of dividing the collection based on publishing decade is further investigated by the correlation between the ranked lists produced by the expanded queries formed from the individual decades. 
We compute the rank correlation between the two ranked lists produced by $EQ_C$ and $EQ_C^i$ in different settings.
In Table~\ref{tab:kendall}, we report the rank correlation (Kendall's $\tau$) between the ranked lists with the two expanded queries.
The results indicate that the correlation between the ranked lists is generally very low, with the highest observed correlation being a trivial $0.130$ for the query \textit{Emigrant}, suggesting that the expanded queries derived from different sub-collections lead to significantly different retrieval results.

To further quantify the similarity in term selection by RLM, we compute the Jaccard similarity between the terms selected by RLM for the considered collections.
Specifically,  we evaluate the Jaccard similarity between any pair of sub-collections $\mathcal{C}_i$ and the entire collection $\mathcal{C}$, as well as between pairs of sub-collections themselves.
For each pair of decades, $\mathcal{C}_i$ and $\mathcal{C}_j$, we calculate the average Jaccard similarity based on the expanded queries generated by RLM for all 25 queries. 
To quantify the variability in term overlap across different topic pairs, the standard deviation of these Jaccard similarities is also reported.
The average Jaccard similarity between the expanded query term sets for all 25 topics is presented in the upper triangle of Table~\ref{tab:similarity}. 
Each cell $(i,j)$ within this triangle displays the mean Jaccard similarity between collections $\mathcal{C}^i$ and $\mathcal{C}^j$.
To provide a measure of dispersion, the standard deviation of these Jaccard similarities is included in parentheses within each cell.

The JS divergence is also computed and presented in the lower half of Table~\ref{tab:similarity} to complement the Jaccard similarity analysis.
While the Jaccard measure focuses solely on the overlap of terms in two expanded query sets, disregarding term weights, JS divergence takes these weights into account and calculates the divergence between a pair of expanded queries from two different sub-collections.
Interestingly, despite variations in specific term compositions (indicated by lower Jaccard values), the average JS divergence across all decade pairs remains relatively constant around $0.5$ with lowest divergence value of $0.48$.
A relatively lower standard deviation across all queries indicates minimal variation among the values, suggesting both the Jaccard similarity and JS divergence for all queries within a decade exhibit consistent patterns. 
This suggests a progressive divergence in the semantic space of query expansions, highlighting the dynamic nature of language and conceptual frameworks over time.
This also indicates a consistent level of semantic separation between the different time periods which varies rarely.


\begin{table}[!t]

\centering

\caption{
\small
Average variations of feedback term distributions obtained by RLM comparing decades of 19th century. Upper triangle (green cells) shows Jaccard similarities of feedback queries generated from the corresponding row and column decades, whereas lower part (red cells) depicts JS divergence between two feedback term distributions. Each cell reports average variation and standard deviation (in brackets) of 25 keyword queries as in Table~\ref{tab:qterms}. Highest Jaccard similarity and lowest JS divergence are bold-faced.
}
\label{tab:similarity}

\begin{adjustbox}{width=1\columnwidth}

\begin{tabular}{@{}lcccccccc@{}}

\toprule

& 1830s & 1840s & 1850s & 1860s & 1870s & 1880s & 1890s & 1831-99\\

\cmidrule{2-9}

\multirow{2}{*}{1830s} 
& 1 
& \cellcolor{magicmint}0.1756 
& \cellcolor{magicmint}0.1718 
& \cellcolor{magicmint}0.1517
& \cellcolor{magicmint}0.1541
& \cellcolor{magicmint}0.1539
& \cellcolor{magicmint}0.1322
& \cellcolor{magicmint}0.1405 \\

& (0)
& \cellcolor{magicmint}(0.1392) 
& \cellcolor{magicmint}(0.1430) 
& \cellcolor{magicmint}(0.1245)
& \cellcolor{magicmint}(0.1275)
& \cellcolor{magicmint}(0.1193)
& \cellcolor{magicmint}(0.1160)
& \cellcolor{magicmint}(0.1220) \\

\multirow{2}{*}{1840s} 
& \cellcolor{melon}0.5249 
& 1
& \cellcolor{magicmint}0.2618 
& \cellcolor{magicmint}0.2541
& \cellcolor{magicmint}0.2465
& \cellcolor{magicmint}0.2383
& \cellcolor{magicmint}0.2245
& \cellcolor{magicmint}0.2257 \\

& \cellcolor{melon}(0.2084)
& (0)
& \cellcolor{magicmint}(0.1422) 
& \cellcolor{magicmint}(0.1314)
& \cellcolor{magicmint}(0.1295)
& \cellcolor{magicmint}(0.1241)
& \cellcolor{magicmint}(0.1361)
& \cellcolor{magicmint}(0.1377)
\\

\multirow{2}{*}{1850s}
& \cellcolor{melon}0.5190 
& \cellcolor{melon}0.5188 
& 1
& \cellcolor{magicmint}0.3214
& \cellcolor{magicmint}0.2967
& \cellcolor{magicmint}0.2801
& \cellcolor{magicmint}0.2717
& \cellcolor{magicmint}0.3166 \\

& \cellcolor{melon}(0.2140) 
& \cellcolor{melon}(0.1954) 
& (0)
& \cellcolor{magicmint}(0.1108)
& \cellcolor{magicmint}(0.1181)
& \cellcolor{magicmint}(0.1047)
& \cellcolor{magicmint}(0.1256)
& \cellcolor{magicmint}(0.1307) \\

\multirow{2}{*}{1860s} 
& \cellcolor{melon}0.5341
& \cellcolor{melon}0.5616
& \cellcolor{melon}0.5469 
& 1
& \cellcolor{magicmint}0.3304
& \cellcolor{magicmint}0.3166
& \cellcolor{magicmint}0.2946
& \cellcolor{magicmint}0.3286 
\\

& \cellcolor{melon}(0.1575) 
& \cellcolor{melon}(0.1434) 
& \cellcolor{melon}(0.1770)
& (0)
& \cellcolor{magicmint}(0.1117)
& \cellcolor{magicmint}(0.1173)
& \cellcolor{magicmint}(0.1214)
& \cellcolor{magicmint}(0.1248) \\

\multirow{2}{*}{1870s} 
& \cellcolor{melon}0.5633
& \cellcolor{melon}0.5403
& \cellcolor{melon}0.5274 
& \cellcolor{melon}0.5493 
& 1
& \cellcolor{magicmint}0.3450
& \cellcolor{magicmint}0.3064
& \cellcolor{magicmint}0.3543 \\

& \cellcolor{melon}(0.1880) 
& \cellcolor{melon}(0.1776) 
& \cellcolor{melon}(0.1963)
& \cellcolor{melon}(0.1766)
& (0)
& \cellcolor{magicmint}(0.0928)
& \cellcolor{magicmint}(0.1198)
& \cellcolor{magicmint}(0.1217) \\

\multirow{2}{*}{1880s} 
& \cellcolor{melon}0.5912
& \cellcolor{melon}0.5443
& \cellcolor{melon}0.5899 
& \cellcolor{melon}0.5267
& \cellcolor{melon}0.5283
& 1
& \cellcolor{magicmint}0.3354
& \cellcolor{magicmint}0.3646 \\

& \cellcolor{melon}(0.2082) 
& \cellcolor{melon}(0.1726) 
& \cellcolor{melon}(0.1064)
& \cellcolor{melon}(0.1971)
& \cellcolor{melon}(0.1975)
& (0)
& \cellcolor{magicmint}(0.1126)
& \cellcolor{magicmint}(0.1190) \\

\multirow{2}{*}{1890s} 
& \cellcolor{melon}0.5234
& \cellcolor{melon}\textbf{0.4817}
& \cellcolor{melon}0.5891 
& \cellcolor{melon}0.5892
& \cellcolor{melon}0.5918
& \cellcolor{melon}0.5732
& 1
& \cellcolor{magicmint}\textbf{0.4491} \\

& \cellcolor{melon}(0.1594) 
& \cellcolor{melon}(\textbf{0.2233}) 
& \cellcolor{melon}(0.1031)
& \cellcolor{melon}(0.1072)
& \cellcolor{melon}(0.1004)
& \cellcolor{melon}(0.1393)
& (0)
& \cellcolor{magicmint}(\textbf{0.1484}) \\

\multirow{2}{*}{1831-99} 
& \cellcolor{melon}0.5483
& \cellcolor{melon}0.5835
& \cellcolor{melon}0.5055 
& \cellcolor{melon}0.5912
& \cellcolor{melon}0.4861
& \cellcolor{melon}0.6133
& \cellcolor{melon}0.5505
& 1 \\

& \cellcolor{melon}(0.1593) 
& \cellcolor{melon}(0.1078) 
& \cellcolor{melon}(0.2121)
& \cellcolor{melon}(0.1005)
& \cellcolor{melon}(0.2309)
& \cellcolor{melon}(0.0085)
& \cellcolor{melon}(0.1704)
& (0)
\\

\bottomrule

\end{tabular}
\end{adjustbox}
\end{table}

\section{Conclusion and future work}\label{sec:conclu}





In this paper, we presented our findings on the study of evolving language usage in 19th century fiction via an information retrieval approach. 
We investigated how the use of words has evolved over time by applying a Relevance-based Language Model, an effective query expansion technique that leverages the statistical occurrences of words. 
Our approach compares the terms selected in one decade with those from another, focusing on the overlap of expansion terms. 
Additionally, we analysed the differences in term importance by computing the divergence in their weights. 

As part of future work, we plan to conduct a comparative study of word usage in 19th century novels and non-fiction texts from the same period, as provided by the British Library Digital Collection. Additionally, we aim to develop an interactive tool that enables humanities researchers to visualise and explore word usage trends in the context of the original documents.

\subsubsection*{\textbf{Acknowledgement}}
This publication is part of a project that has received funding from (i) the European Research Council (ERC) under the Horizon 2020
research and innovation program (Grant agreement No. 884951); (ii) Science Foundation Ireland (SFI) to the Insight Centre for Data
Analytics under grant No 12/RC/2289\_P2.

\bibliographystyle{acm-reference-format}
\balance
\bibliography{main}

\end{document}